\documentclass[fleqn,usenatbib]{mnras}

\usepackage{newtxtext,newtxmath}

\usepackage[T1]{fontenc}
\usepackage{ae,aecompl}

\usepackage{graphicx}	
\usepackage{amsmath}	
\usepackage{amssymb}	
\usepackage{gensymb}
\usepackage{subfigure}

\title[Gamma-rays from SS433]{Gamma-rays from SS433: evidence for periodicity}

\author[K. Rasul et al.]{
Kajwan Rasul,$^{1}$
Paula M. Chadwick,$^{1}$\thanks{E-mail: p.m.chadwick@durham.ac.uk (PMC)}
Jamie A. Graham$^{1}$ and Anthony M. Brown$^{1}$
\\
$^{1}$Centre for Advanced Instrumentation, Dept. of Physics, Durham University, Durham DH1 3LE, UK\\
}

\date{Accepted 2019 February 21. Received 2019 February 21; in original form 2018 October 4.}

\pubyear{2018}

\begin{document}
\label{firstpage}
\pagerange{\pageref{firstpage}--\pageref{lastpage}}
\maketitle

\begin{abstract}
In this paper we present our study of the gamma-ray emission from the microquasar SS433. Integrating over 9 years of \textit{Fermi}-LAT \textsc{pass8} data, we detect SS433 with a significance of $\sim$~13~$\sigma$ in the 200 to 500 MeV photon energy range, with evidence for an extension in the direction of the w1 X-ray `hotspot`. A temporal analysis reveals evidence for modulation of SS433's gamma-ray emission with the precession period of its relativistic jet. This suggests that at least some of SS433's gamma-ray emission originates close to the object rather than from the jet termination regions.
\end{abstract}




\section{Introduction}
SS433 was the first detected microquasar and even today is regarded as the only known supercritical accretor in the Milky Way. The binary is well-studied and consists of a compact object, thought to be a 10-20 $M_{\odot}$ black hole \citep{gie02}, and an approximately 11~$M_{\odot}$ A3-7I star \citep{hil04}. The orbital period of the system is 13.1 days \citep{hil04}. SS433 was the first X-ray binary system shown to have highly energetic jets \citep{abe79,fab79}. The two jets are being ejected from the system at a velocity of 0.26$c$, and precess every 162.4 days \citep{mar89,eik01}.  SS433 lies in the centre of the supernova remnant W50, and its impact on the surrounding region is clearly observed in radio maps \citep[e.g.][]{dub98,bel11}. Radio observations have revealed two distinct jet states, similar to those observed in the case of Cygnus X-3 \citep{joh81,bon86}. One of these states is strongly related to major radio flares, which are also regularly detected from Cygnus X-3.

Gamma-ray emission from the direction of SS433 was reported by \citet{bor15}, using five years of data obtained with the \textit{Fermi} Large Area Telescope. They reported a  7.3 $\sigma$ detection with a 99.9 \% confidence level position accuracy; their investigation did not lead to any detection of the orbital or precession period of the binary. There was no emission observed from the object above  $\sim800$~MeV. While very high energy gamma-ray emission from the object has been proposed (see, e.g., \citep[see, e.g.,][]{aha98,bos06,rey08,bor09}, a recent study by the MAGIC and H.E.S.S. collaborations shows no evidence for any emission from the object above 300 GeV \citep{ahn18}, suggesting that little of the jets' kinetic power is transferred to $\gamma$-ray emission. However, recent observations with the HAWC telescope have revealed evidence for emission at energies above 20~TeV from two sites coincident with the X-ray `hotspots' w1 and e1 associated with the jet termination regions, some 40~parsecs from the central source \citep{abe18}. There is no evidence for emission from the central source.

Here we present our analysis of \textit{Fermi}-LAT \textsc{pass8} data from the region of SS433. These data provide both improved PSF classes, over an extended energy range, and nearly double the exposure compared to previous studies. In Section 2, we describe the \textit{Fermi}-LAT data selected for analysis and describe the analysis undertaken. In Section 3, we show the results of our analysis, including a detailed temporal search of the gamma-ray emission modulation associated with either the orbital and precession periods, including any spectral changes. In Section 4 we discuss the implications of our results. 

\section{\textit{Fermi} Analysis}
\label{sect:fermi}
Our analysis of SS433 is based on 9 years of \textit{Fermi}-LAT data from the 4th of August 2008 to the 9th of August 2017 (MET: 239557417-523943505). All \textsc{pass8 source}  ``FRONT+BACK" $\gamma$-ray events, in the 0.1 - 1 GeV energy range, within a 15$\degree$ radius of interest (ROI) centred on the position of SS433, (RA, Dec = 287.957$\degree$, 4.983$\degree$), were considered. As required for \textsc{pass8} data analysis, a 90$\degree$ zenith cut was applied to the data and the good time intervals were created by excluding periods where \textit{Fermi} passed through the `South Atlantic Anomaly', or where on board spacecraft incidents have the potential to affect the quality of the data. This was achieved by applying a filter expression of `\textsc{data\_qual}$>0$ \&\& \textsc{lat\_config}$==1$' to the data that satisfied the zenith angle criterion. 

Throughout our analysis, version \textsc{v10r0p5} of the \textit{Fermi Science Tools} and version 0.12.0 of \textit{fermipy} was used in conjunction with the P8R2\textunderscore SOURCE\textunderscore V6 instrument response function. For the likelihood analysis, a model consisting of diffuse, extended and point sources of $\gamma$-rays was employed. For the diffuse emission the most recent Galactic and isotropic diffuse models were used, gll\textunderscore iem\textunderscore v06.fit and iso\textunderscore P8R2 \textunderscore SOURCE\textunderscore V6\textunderscore v06.txt respectively. The extended sources in our model were the supernova remnants W51C and W44, which were described by the spatial map provided by the \textit{Fermi}-LAT collaboration. The $\gamma$-ray point source population of our model was initially seeded by the third \textit{Fermi} catalog \citep[3FGL;][]{3fgl}, taking the position and spectral shape of all 3FGL point sources within 25$\degree$ of SS433. \par

The first step of our analysis was a \textsc{binned} likelihood analysis, using the \textit{fermipy} \textsc{optimize} routine. From this initial optimization, all sources with a test statistic\footnote{The test statistic, TS, is defined as twice the difference between the log-likelihood of different models, 2(log $\mathcal{L}_{1} -$ log $\mathcal{L}_{0})$, where $\mathcal{L}_{1}$ and $\mathcal{L}_{0}$ are defined as the maximum likelihood with and without the source in question \cite{like}} TS$<2$ or a predicted number of photons $Npred<4$ were removed from the model\footnote{Some of the TS$<2$ objects removed are 3FGL objects, as some of these sources are variable and are not significant in the larger dataset analysed here.}. Thereafter a point source, modelled with a power law spectral shape of the form $dN / dE=A \times E^{-\Gamma}$, was added to the centre of the ROI, where SS433 is expected to be, and an additional \textsc{optimize} routine was undertaken. 

The best-fit model resulting from this initial model was then used, in conjunction with the \textit{Fermi} science tool \textsc{gttsmap}, to create a $21\degree\times21\degree$ TS map centred on SS443. This TS map was used to reveal additional point sources of gamma-rays that were not accounted for in our initial model by identifying excesses in the TS map with TS$>25$. These additional point sources were modelled with a power law, and the RA and Dec of each new source taken from the position of the maximum excess. Having accounted for all sources of gamma-rays in our dataset, the position of SS433's gamma-ray emission was refined using the \textit{fermipy} \textsc{localize} function which removes the source from the model, makes a TS map and adds the source to the points which maximise the log-likelihood of the model. Finally, the power law spectral model for the SS433 assumption was relaxed, and we fitted SS433 with both log-parabola and broken power-law spectral models, to see if this led to a better fit for the model. 

\section{Results}
\subsection{Detection Significance}

Using the approach described above, SS433 is detected with a test statistic of TS $= 173$ corresponding to a $\sim$13 $\sigma$ detection. The final model consists of 72 sources which include 2 extended sources (SNR W44 and W51C), 36 3FGL point sources and 33 new point sources that are not in the 3FGL. Table 1 shows all new point sources discovered within 5$\degree$ of SS433. \par 

\begin{table*}
	\centering
	\label{tab:model}
   
    \begin{tabular}{lccccccc}
    		\hline
    Source Name & RA & Dec & Pos. Err & Flux  & $\Gamma$ & TS & Offset \\  		
       & deg. & deg. & deg. & 10$^{-8}$ cm$^{-2}$ s$^{-1}$ &  &  & deg. \\  		\hline

PS J1905.2+0351 & 286.30 & +3.86 & 0.12 & 5.92$\pm$0.59 & 2.51$\pm$0.12 & 145.7 & 2.0 \\
PS J1918.3+0640 & 289.60 & +6.68 & 0.10 & 5.08$\pm$0.47 & 2.73$\pm$0.12 & 160.0 & 2.4 \\
PS J1910.5+0736 & 287.63 & +7.61 & 0.11 & 5.45$\pm$0.58 & 2.29$\pm$0.13 & 129.6 & 2.6 \\
PS J1859.2+0559 & 284.80 & +5.99 & 0.09 & 8.61$\pm$0.61 & 2.38$\pm$0.09 & 291.7 & 3.3 \\
PS J1907.5+0920 & 286.98 & +9.33 & 0.06 & 2.39$\pm$0.25 & 1.44$\pm$0.85 & 94.3 & 4.4 \\
PS J1854.6+0311 & 283.65 & +3.20 & 0.11 & 8.03$\pm$0.64 & 2.18$\pm$0.09 & 271.2 & 4.6 \\
PS J1852.2+0533 & 283.07 & +5.56 & 0.18 & 4.97$\pm$0.55 & 2.58$\pm$0.14 & 114.0 & 4.9 \\
\hline

    \end{tabular}
   \caption{All significant new sources added to the model within 5 degrees of SS433. $\Gamma$ is the best-fit power-law index for the emission.}
\end{table*}

\subsection{Source Localization}
\begin{figure}
	\includegraphics[width=\columnwidth]{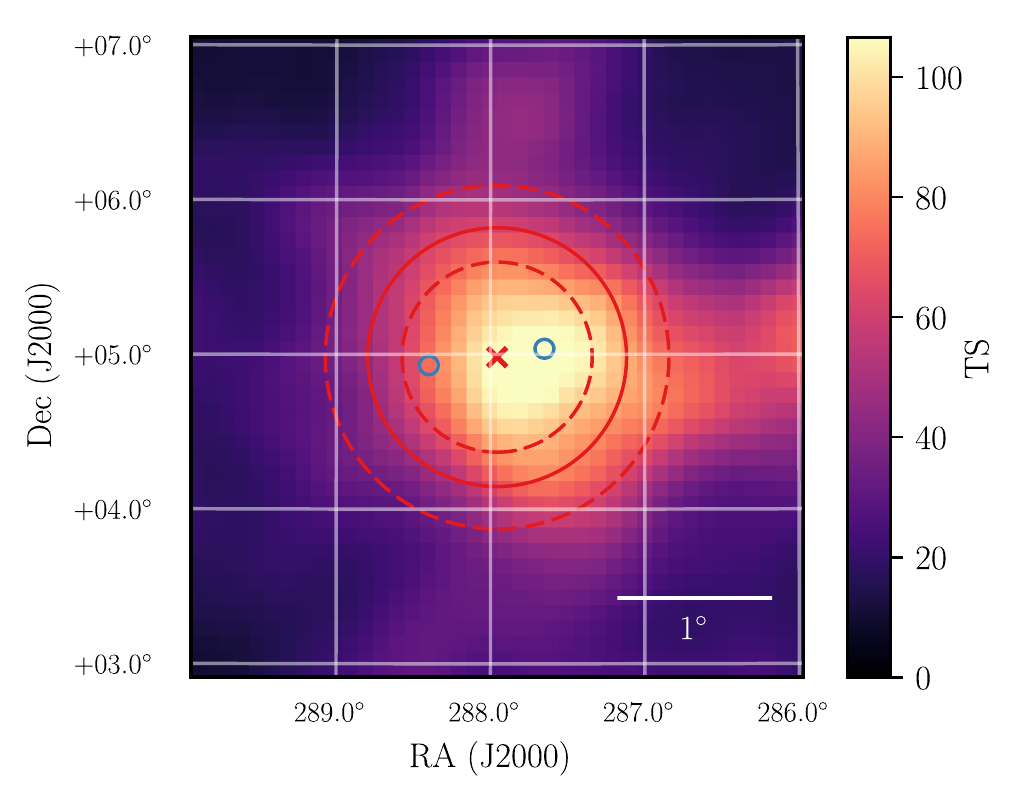}
    \caption{$TS$ map centred on the position of SS433. The solid red circle indicates the optimised extension, with dotted red circles representing the 1$\sigma$ uncertainty on the extension. The red cross represents the position of SS433 in this analysis, and blue circles represent the positions of the West and East lobes of SS433 as defined by \citet{abe18}.}
    \label{fig:tsmap}
\end{figure}

The best-fit position of the gamma-ray emission is (RA, Dec = 287.806$\degree$, 4.871$\degree$), with a 95\% positional uncertainty of 0.240$\degree$. We note that the best-fit position is 0.188$\degree$ from SS433, though this is a significant improvement on the position found in \citet{bor15} who reported an offset of 0.41$\degree$. A 15$\degree \times 15\degree$ TS map, centred on SS433, is shown in Figure \ref{fig:tsmap}, which clearly shows a significance excess positionally coincident with SS433. The TS map also reveals some evidence for extended gamma-ray emission roughly coincident with SS433's radio lobes, especially towards the w1 lobe as studied in \citet{abe18}. A check for extended emission finds a TS$_\text{ext}$ of 31.1 (as defined in \citet{ack18}) at the catalog position of SS433 with a best-fit extension of (0.84$\pm$0.27)$^\circ$.

We note that a recent paper (\citet{xin18}) found evidence for emission from the w1 lobe using a similar dataset to the one which we have analysed. \citet{xin18} use the FL8Y catalogue, for which (as the authors point out) the Galactic and extragalactic diffuse models have not been updated. This difference in analysis is likely responsible for the different morphologies observed.

\subsection{Spectrum}
SS433's spectrum, in the 0.1--1GeV energy range, can be seen in Figure 2. The best-fit to the spectrum is a log-parabola function: \par
\begin{equation}
\frac{dN}{dE} = (7.73 \pm 1.08) \times 10 ^{-13} \Big(\frac{E}{10^{3}}\Big)^{-4.87 + 1.0 log(E/10^{3})} 
\end{equation}

in the 0.1--1~GeV energy range and is preferred over a power law by $\sim4\sigma$. \citet{bor15} also found that the spectrum was best-fitted by a log-parabola, the parameters of which are compatible with our fit, and we also note the presence of a maximum around 250~MeV. \citet{bor15} commented on the lack of emission above 800~MeV; the measurement reported here shows no evidence for emission above 500~MeV. \par

\begin{figure}
	\includegraphics[width=\columnwidth]{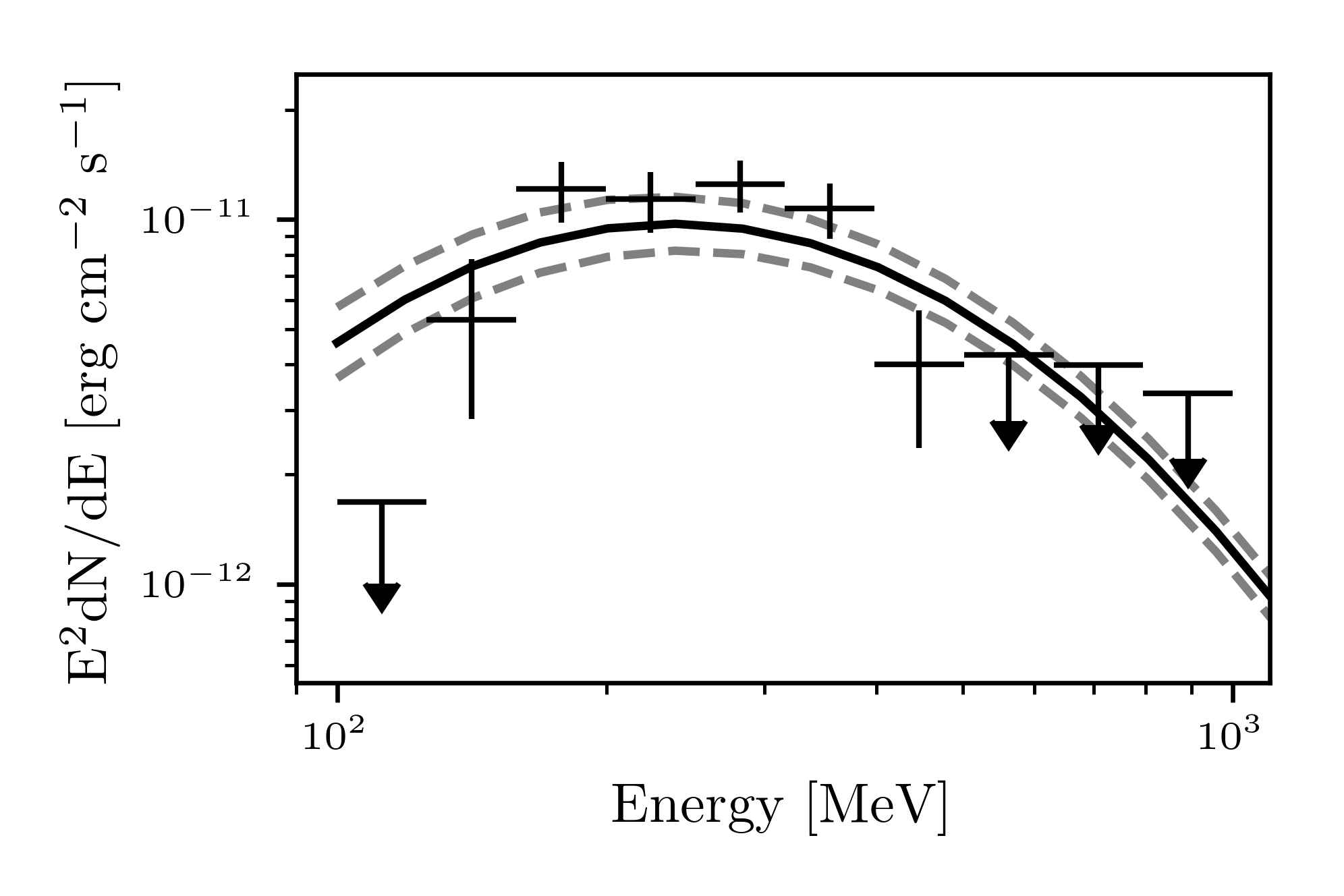}
    \caption{Spectral Energy Distribution of SS433 in the 0.1--1 GeV range. The best-fit log-parabola function is shown, with the dotted lines indicating the uncertainties on the functional fit. For energy bins with a TS$>25$, the flux is shown, along with the $1\sigma$ statistical uncertainties. Upper limits are 2$\sigma$ limits.}
    \label{fig:sed}
\end{figure}

\subsection{Periodicity}
Since, in the case of SS433, the \textit{Fermi}-LAT integration times necessary for a significant detection of the object are longer than the orbital and precessional periods, any periodicity is smeared out in the simple flux vs. time light curve. The only way to detect the periodicity is therefore to fold the data with the \textit{a priori} periods of the object as defined by observations at other wavelengths. We note that \citet{xin18} searched for variability, but their analysis was very different from ours. They searched for variability from the w1 lobe rather from the position of SS433, considered events only above 300~MeV (we find a cut-off at around 500 MeV), and binned the data in 30-day bins, which would be too long to detect any orbital periodicity.

Accordingly, to investigate periodic variations in SS433's gamma-ray properties, we used the \textit{Fermi Science Tools}, \textsc{gtpphase}, which assigns a phase to each photon event in the data if given an ephemeris and a period. This was used to split the data up into 10 phase bins for both the orbital and precession period. The analysis described in Section \ref{sect:fermi} was applied to each of these bins to make the orbital and precession light curves. The ephemeris applied for the orbital analysis was that of \citet{kat83}. The orbital period from \citet{kat83} produces a phase error of 0.2 days when rolled forwards to the epoch of the observations which we analysed, compared to 0.55 days for that of \citet{and83}. For the precession analysis, we have applied the ephemeris of \citet{eik01}, which provides a phase error of 0.11~d, compared with a phase error of 0.68~d for the \citet{and83} ephemeris; the other ephemerides available do not enable a phase error on the precession period to be calculated. The parameters of the ephemerides applied are provided in Table 2; the observation epoch was set to 55000 MJD in both cases. The two resulting light curves are shown in Figures~\ref{fig:Lightcurves} (a) and \ref{fig:Lightcurves} (c).

\begin{table*}
	\centering
	\label{tab:ephem}
	\begin{tabular}{lcccc} 
		\hline
		& Period (d) & Period Error (d) & Zero Phase (JD) & Phase Error (d)\\
		\hline
		Orbital & 13.0682 & 0.0035 & 2443551.93 & 0.20\\
        Precession & 162.375 & 0.011 & 2443563.23  & 0.11 \\ 
		\hline
	\end{tabular}
   \caption{Orbital and precession ephemerides used in the analysis. The orbital ephemeris is from Katz et al. (1983) and the precession ephemeris from Eikenberry et al. (2001). The phase error is the accumulated error from the date of the ephemeris to the time of the observations.}
\end{table*}

Given the apparent variability shown in the lightcurves, we look to assess the following models of periodicity:
\begin{itemize}
    \item A) A constant flux $N_0$ with no dependence on phase $\phi$:
    \begin{equation*}
    F(\phi) = N_0.
    \end{equation*}
    \item B) A constant flux with an additive sinusoid of amplitude $C$ with a minimum at $\phi=-0.25$:
    \begin{equation*}
    F(\phi) = N_0 + C\sin(2\pi\phi).
    \end{equation*}
    \item C) A constant flux with an additive sinusoid and an optimised phase offset $\phi_0$:
    \begin{equation*}
    F(\phi) = N_0 + C\sin(2\pi(\phi + \phi_0)).
    \end{equation*}
\end{itemize}

To assess the significance of the alternative hypotheses, models B and C, in comparison to our null hypothesis model A, we perform a Monte Carlo simulation. For the case of both orbital and precessional periodicity, our approach is the same. We generate $10^5$ datasets assuming the null hypothesis of constant flux. The flux of each phase bin is sampled from the distribution $\mathcal{N}(\mu, \sigma^2)$, where $\mu$ is the best-fit constant flux and $\sigma$ is the standard deviation in our measured flux. Similarly, the uncertainty on each generated flux is sampled from $\mathcal{N}(\mu_\text{unc}, \sigma_{\text{unc}}^2)$ where $\mu_\text{unc}$ and $\sigma_\text{unc}$ are the equivalent mean and standard deviation in our measured statistical flux uncertainty. Each generated dataset is fit by minimising the $\chi^2$ value obtained with models B and C, and the fraction of events for which the $\chi^2$ value exceeded the $\chi^2$ for our measured data determined the significance of periodicity. This effectively represents a Monte Carlo approach to the $\chi^2$ method introduced by \citet{lea83} for folded lightcurves. This method was validated by application to 12 background point sources and found no evidence of variability.

\subsubsection{Orbital Variability}

An additive sinusoid with a fixed minimum at $\phi=0.0$ (model B) is attained in 7.15\% of simulated datasets. While allowing the phase of the sinusoid to vary (model C), this is achieved in only 5.30\% of simulated datasets. This corresponds to approximately 2$\sigma$ evidence of sinusoidal variability in the orbital lightcurve when compared to an assumption of constant flux (model A). The optimised model parameters are in this case:
\begin{equation}
    F(\phi) = (1.07 + 0.18\sin(2\pi(\phi + 0.81))) \times 10^{-8}\,\text{ph}\,\text{cm}^{-2}\,\text{s}^{-1}.
\end{equation}

\subsubsection{Precessional Variability}
We apply these methods to the precessional lightcurve, but without considering model B as we have no \emph{a priori} evidence to suggest that the emission should be strongest at $\phi=0.0$ in the case of precession. In the case of model C, we find stronger evidence of temporally-correlated emission than in the orbital lightcurve, with only 1.4\% of simulated datasets with a $\chi^2$ value smaller than the observed data, a result significant at the 2.45 $\sigma$ level. If the phase parameter of the sinusoid is profiled out as a free parameter, this result improves to the 2.9 $\sigma$ level.

\begin{figure}
\includegraphics[width=\linewidth]{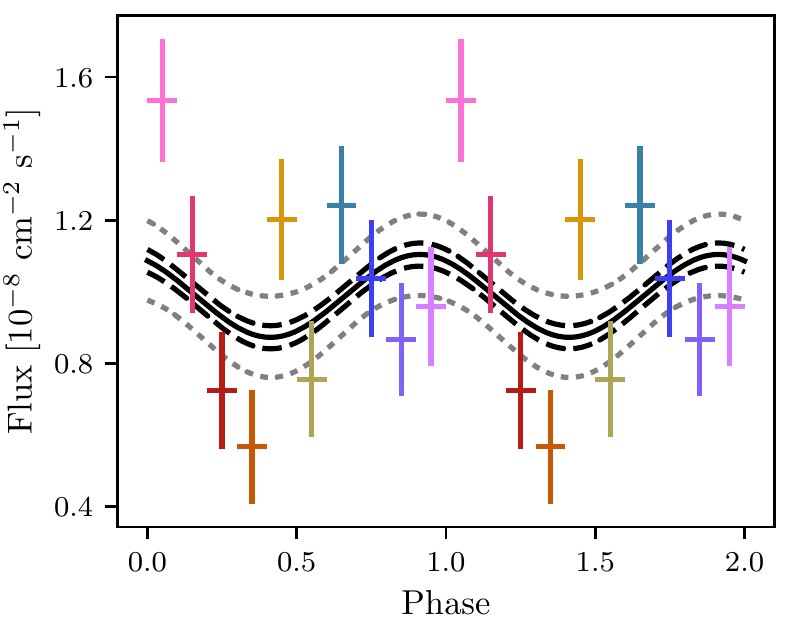}
\caption{Reconstructed flux from our fit to the data. The solid black line corresponds to the median flux as a function of phase, and the dashed and dotted lines correspond to the 1 and 2$\sigma$ uncertainty regions as evaluated as percentiles of the posterior's sampling when evaluated for flux as a function of phase.}
\label{fig:mcmc_precess}
\end{figure}

\begin{figure*}
\includegraphics[width=\linewidth]{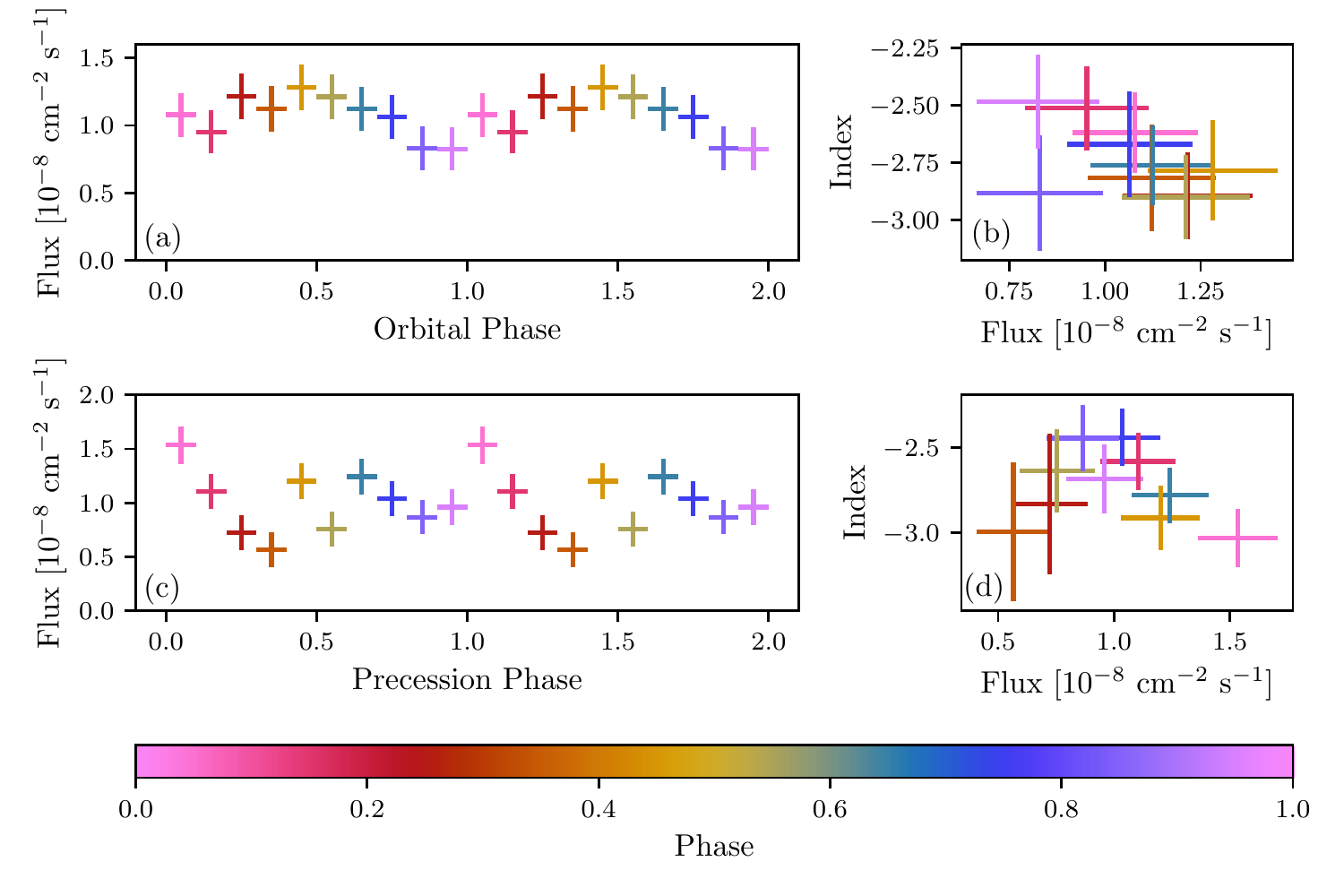}
\caption{(a) Orbital light curve; two orbital periods are shown for clarity. Bins are coloured to match their phase. (b) Spectral index vs flux for each orbital phase bin. (c) Precessional light curve. (d) Spectral index vs flux for each precessional phase bin.}
\label{fig:Lightcurves}
\end{figure*}

In order to estimate the uncertainties on the best-fit parameters, in-depth parameter estimation was performed using an MCMC maximum likelihood fit to the data using the python package \texttt{emcee} \citep{for13}. This allowed us to place confidence regions on the flux as is plotted in Figure~\ref{fig:mcmc_precess}. The optimised parameters in the case of the precession period are:

\begin{equation}
    F(\phi) = (0.99 + 0.14\sin(2\pi(\phi + 0.84))) \times 10^{-8}\,\text{ph}\,\text{cm}^{-2}\,\text{s}^{-1}.
\end{equation}

\section{Discussion}

We have noted a similar spectral cut-off to that of \citet{bor15}. In their analysis, the best-fit to this spectrum was provided by a model in which gamma-ray emission is produced by the decay of neutral pions produced by proton-proton collisions. Such a scenario requires a population of accelerated protons, which could naturally be provided by the relativistic jet. However, given the strong cut-off in the spectrum, this emission mechanism would need to be rather inefficient, which also agrees with the conclusions of the MAGIC and H.E.S.S. collaborations in relation to the lack of VHE emission, at least from the central object \citep{ahn18}; as already noted, there is evidence for emission at much higher energies from the lobes \citep{abe18}. 

\citet{bor15} concluded that the likely site for this inefficient process would be close to the compact object so that the protons could interact with cold jet material or the disc wind. In this case, where the emission site is close to the binary system, one would expect to see both the precession period and the orbital period imprinted on the gamma-ray data \citep{rey08}, which was not apparent in the smaller dataset which they analysed. They therefore concluded that the likely site for the gamma-ray production was the jet termination regions. However, we now find some evidence for periodicity, at least in the case of the precession period, which is seen in the data at the $\sim3~\sigma$ level of significance. \citet{rey08} also suggest that the gamma-ray emission should be weakest around precession phase 0.5 and strongest in the range of precession phases between 0.91 and 0.09, i.e. around zero phase, which is compatible with the precessional light curve we obtained.

Recent work by \citet{mol18} on a general model for high-mass microquasar jets notes that strong orbital modulation may be expected in gamma-ray emission which is placed further from the compact object, at the point where a helical jet (known to exist in SS433) bends from its initial trajectory. The orbital modulation predicted by the model would have a maximum at phase 0.5, which is also the case for our orbital light curve (such as it is). However, their model is leptonic in nature, with the inverse Compton emission maximising around 10~GeV. This is not compatible with the cut-off that is observed in our spectrum, but a more detailed multiwavelength spectral analysis and modelling would be necessary to distinguish conclusively between models. Finally, we note that the HAWC result favours the existence of accelerated electrons rather than protons in the lobes \citep{abe18}.

\section{Conclusions}

We have detected SS433 at a TS level of 173, corresponding to a significance of $\sim$13 $\sigma$, in 9 years of \textit{Fermi}-LAT data. The spectrum is best fit by a log-parabola; there is no evidence for emission above 500 MeV. The centroid of the \textit{Fermi} emission is offset by 0.188$\degree$ from the nominal position of SS433, but still within the 95\% positional uncertainty. There is evidence at the $\sim 3~\sigma$ level for modulation of the gamma-ray emission with the precession period of the jet, but there is no significant evidence for orbital modulation of the emission. These results suggest that at least some of SS433's gamma-ray emission originates close to the base of the jet.

HE gamma-ray emission is observed from three other likely jet-powered objects:  V404~Cygni \citep[][]{loh16,pia17}, Cygnus~X-1 \citep{mal13,zan16} and  Cygnus~X-3 \citep{abd09,bul12,cor12}. Emission is generally seen during outburst or X-ray hard states, although other prominent microquasars, such as GRS1915+105 and GX339+4, have not been detected in HE gamma rays thus far, even during high or outburst states \citep{bog13}. The brightest of the detected objects, Cygnus X-3, shows some evidence for persistent emission \citep{bog13} and the emission is strongly orbitally-modulated during flares \citep{zdz18}. There is a hint of orbital modulation in the HE emission from Cygnus X-1, with most of the emission occurring at superior conjunction \citep{zan16}, while the detection of V404~Cygni at $\sim4~\sigma$ is too weak to enable any orbital modulation to be detected, should it be present. To these we now add SS433, with evidence for periodicity at the precession period, but not the orbital period. It is likely that the presence or otherwise of any periodicity within the HE gamma ray data from these objects is more strongly related to the statistical significance of the detection than any physics, since the best evidence for periodicity, on the basis of this rather small sample, comes from the most strongly-detected objects.

It is not possible on the basis of these results to say conclusively whether hadronic or electronic processes dominate in SS433. Further gamma-ray observations are required with more sensitive instruments, such as the Cherenkov Telescope Array \citep{ach13}. The observation of neutrinos from SS433 would provide clear evidence for hadronic processes. However, assuming that the cut-off we see at around 500 MeV is not due to photon-photon absorption, there would be little, if any, neutrino signal at the 10 GeV energy threshold of the IceCube DeepCore low energy subdetector \citep{abb12}.

\section*{Acknowledgements}

This paper makes use of publically-available \emph{Fermi}-LAT data provided online by the \href{http://fermi.gsfc.nasa.gov/ssc/}{Fermi Science Support Center}. We acknowledge the excellent quality of the data and analysis tools provided by the \textit{Fermi}-LAT collaboration. JAG is supported by the STFC, grant reference ST/N50404X/1, and PMC and AMB acknowledge support from STFC grant ST/P000541/1. 







\bsp	
\label{lastpage}
\end{document}